\documentclass[conference]{IEEEtran}
\IEEEoverridecommandlockouts
\usepackage{cite}
\usepackage{amsmath,amssymb,amsfonts,amsthm}
\usepackage{comment}
\usepackage{algorithmic}
\usepackage{graphicx}
\usepackage{textcomp}
\usepackage{xcolor}
\usepackage{bbm}
\usepackage{tikz, pgfplots, subcaption}
\usepackage{bbold}

\usetikzlibrary{arrows.meta}
\usetikzlibrary{datavisualization}
\usetikzlibrary{external}

\newtheorem{theorem}{Theorem}%[section]
\newtheorem{proposition}[theorem]{Proposition}%[section]

\newtheorem{assumption}{Assumption} %[section]
\newtheorem{remark}{Remark}
\newcommand{\Var}{\mathrm{Var}}

\newcommand{\mc}{\mathcal}

\newcommand{\tcb}{\textcolor{black}}

\pgfplotsset{compat=1.18}

\begin{document}

%Title from Anton
\title{\vspace{20pt}Continuous-Time Distributed Learning\\for Collective Wisdom Maximization

\thanks{
 Luka Baković, Giacomo Como and Emma Tegling are with the Department of Automatic Control, Lund University, Lund, Sweden. They are members of the ELLIIT Strategic Research Area at Lund University.   
 Email: \{{\tt\small{luka.bakovic,giacomo.como, emma.tegling}\}@control.lth.se}.  
}

\thanks{Giacomo Como and Fabio Fagnani are with the Department of Mathematical Sciences and Anton Proskurnikov is with the Department of Electronics and Telecommunications, both at Politecnico di Torino, Torino, Italy.
 Email: \{{\tt\small{giacomo.como, fabio.fagnani}\}@polito.it; anton.p.1982@ieee.org}.  
}

\thanks{This work was partially supported by ELLIIT and the Wallenberg AI, Autonomous Systems and Software Program (WASP) funded by the Knut and Alice Wallenberg Foundation. G. Como is supported by the Italian Ministry of University and Research under the PRIN project ‘‘Extracting essential information and dynamics from complex networks’’, grant  no. 2022MBC2EZ.
A. Proskurnikov is supported by the 2022K8EZBW ``Higher-order interactions in social
dynamics with application to monetary networks" project -- funded by European Union -- Next Generation EU  within the PRIN 2022 program (D.D. 104 -- 02/02/2022 Ministero dell’Università e della Ricerca). This manuscript reflects only the authors’ views, and the Ministry cannot be considered responsible for them.}
}

\author{ {Luka Baković, Giacomo Como, Fabio Fagnani, Anton Proskurnikov, Emma Tegling} 
}

\maketitle

\begin{abstract}
Motivated by the well established idea that collective wisdom is greater than that of an individual, we propose a novel learning dynamics as a sort of companion to the Abelson model of opinion dynamics. Agents are assumed to make independent guesses about the true state of the world after which they engage in opinion exchange leading to consensus. We investigate the problem of finding the optimal parameters for this exchange, e.g. those that minimize the variance of the consensus value. Specifically, the parameter we examine is susceptibility to opinion change. We propose a dynamics for distributed learning of the optimal parameters and analytically show that it converges for all relevant initial conditions by linking to well established results from consensus theory. Lastly, a numerical example provides intuition on both system behavior and our proof methods.
\end{abstract}

%\begin{IEEEkeywords}
%\end{IEEEkeywords}

% \section{Introduction guidelines}

% \begin{itemize}
%     \item\textcolor{blue}{A deep extension of Moreau's results (and more rigorous): Lin, Francis, Maggiore, State Agreement for Continuous‐Time Coupled Nonlinear Systems,SICON 2007.}
% \end{itemize}

\section{Introduction}

The idea of considering a multitude of opinions to form an accurate description of a phenomenon or come up with a solution to a problem is rather commonplace in today's society. This concept permeates everything from the very practical elements of society, as the principle underlying democratical pluralism, to the folklore, where expressions such as \textit{two heads are better than one} are commonplace. Scientific inquiry into this question is considered to have started with Galton's \textit{Vox Populi} \cite{Galton1907-uj} where it was discovered that the median guess on a numerical value taken from a large population is surprisingly close to the ground truth. One could attempt to write this off as a simple experiment demonstrating that sampling a large number of independent measurements increases the accuracy, but the phenomenon itself is much deeper than that. Namely, human opinions are not completely independent of one another but instead evolve and depend on the opinions of those one comes in contact with. The study \cite{Becker2017-ki} demonstrates that not only do more complicated effects come into play, they can also be harmful to the overall wisdom the group displays. Specifically, they show that social power plays a crucial role. This makes sense, as influential individuals are by definition those who have a greater say or whose opinion matters most. Mathematically, this is closely related to the concept of centrality studied in, for instance, \cite{Katz1953} and \cite{Bonachich}. Certain nodes in graphs enjoy privileged positions due to either having a large number of connections or being connected to the right nodes. Their influence on certain dynamical systems defined over those graphs is then proportional to their centrality. The two concepts have been analyzed in parallel ever since the beginnings of the field of opinion dynamics, but most notably in Friedkin's work \cite{friedkin1999influence}. 
A complementary concept present in, amongst others, Abelson's model \cite{abelson1967mathematical} and Taylor's model \cite{Taylor} is that of an individual's susceptibility to opinion change or self confidence. Instead of just relying on the graph centralities, these models also consider the fact that some individuals are less likely to listen to others in general - regardless of their centrality. Situations where these parameters are dynamic have also been considered, some examples being \cite{Jia2015} where the social influence graph evolves, or \cite{Wang2023} in which the authors concatenate models with differing levels of stubbornness. Finally, \cite{Amelkin2017-ar} considers a nonlinear model dealing with polar opinion evolution, where one of the key components are the dynamical susceptibilities to persuasion which are functions of the agents' current beliefs.

We begin our analysis by considering a group of agents taking independent guesses on the value of some ground truth. We then suppose that the agents partake in an opinion exchange according to the Abelson model. The first question we then pose is related to the model parameters governing the exchange -- how openly should the agents communicate if they want to maximize the "good" effect of crowd wisdom? Mathematically speaking, which susceptibility parameters result in the least possible variance of the group's final guess from the ground truth. After answering this question, we move on to proposing a dynamical system defined over the same set of agents which converges to this optimal parameter set. Our work draws inspiration from some of the authors' previous work~\cite{Como2022-yd}, which examines the optimal self-confidence values in the discrete scenario of French-DeGroot~\cite{degroot1974reaching} by considering it as a game and studying the Nash equilibria. A similar setup is considered in \cite{Golub2010} where the possibility of growing communities is also investigated.

\subsection{Contributions and paper outline}

This paper proposes a novel family of susceptibility adaptation dynamics for the Abelson model. Assuming that agents connected in a graph make independent guesses on the value of a ground truth, we calculate the optimal set of susceptibilities, i.e. the one leading to a crowd estimate that has the smallest possible variance from the ground truth. The dynamical system is then defined as one where each agent moves away from the derivative of a local utility function. The set of fixed points of the system is shown to be exactly the set of optimal susceptibilities, and the system is shown to converge to a point in the set for all relevant initial conditions. Finally, a numerical example provides intuition both on the problem and the proof methods used.

The model is introduced in Section~\ref{sec:model}, along with the main result of the paper. The whole of Section~\ref{sec:analysis} is dedicated to analytically proving all of the building blocks of the main result. A numerical example is presented in Section~\ref{sec:num}. Finally, conclusions and future work are discussed in Section~\ref{sec:conc}.

%\newpage
\section{Problem setup} \label{sec:model}
\subsection{Notation} \label{sec:notation}
Vectors are considered as columns in $\mathbb R^n$. The symbol~$\mathbb{1}$ denotes the vector of all ones and its dimension is implied by the context. When applying inequalities to vectors, they are interpreted element-wise. For example, writing $\mathbf z > 0$ indicates that every component of the vector $z$ is positive.
For a vector $\mathbf z \in \mathbb R^n$, the symbol $[\mathbf z]$ denotes a $n \times n$ diagonal matrix with diagonal $z_1, ..., z_n$. For a function $f : \mathbb R \rightarrow \mathbb R^n$, the symbol $f(\infty)$ is used to denote $\lim_{t \rightarrow +\infty} f(t)$ whenever the limit exists. Summation indices are assumed to run over the set $\{1, ..., n\}$ unless specified otherwise.
% The symbol $\mathcal N_i$ denotes the set of neighbors of vertex $i$ including itself, where the underlying graph is implied by the context.

\subsection{The Model}
Consider a social network modeled as a finite directed weighted graph ${\mathcal G=(\mathcal V,\mathcal E,W)}$, whose node set ${\mathcal V = \left \{ 1, ..., n \right \}}$ represents agents that are connected by a set of directed links ${\mathcal E\subseteq\mathcal V\times \mathcal V}$. Every link ${(i,j) \in \mathcal E}$ has a weight $W_{ij}>0$ representing the strength of the direct social influence exerted by agent $j$ on agent $i$ and, by convention, we set $W_{ij}=0$ for every ${(i,j) \in \mathcal V\times\mathcal V\setminus\mathcal E}$. 
Given a matrix $W$, let $\mathcal{L}[W]=[W\mathbb{1}] - W$ denote the corresponding (weighted) Laplacian associated with $W$.

\begin{assumption} \label{Gcon}
    The graph $\mc G$ is strongly connected.
\end{assumption}
\tcb{Assumption~\ref{Gcon} is equivalent to the irreducibility of $W$.} 
A classical result from Perron-Frobenius theory~\cite{FB-LNS} states that under such \tcb{an} assumption there exists a unique left eigenvector $\mu$ in $\mathbb R^n$ of the Laplacian such that 
\begin{align}\label{eq.mu}
    \mathcal{L}[W]^\top \mu          = 0, \quad
    \mathbb{1}^{\top} \mu = 1,      \quad 
    \mu             > 0.
\end{align}
The elements of $\mu$ are also known to represent the eigenvector centrality. 

Every agent ${i \in \mathcal V}$ is characterized by a scalar opinion ${x_i \in \mathbb R}$ and a positive parameter ${z_i \in (0, \infty)}$ representing her susceptibility to opinion change. 
%\textcolor{red}{TODO: It's not fully clear why we call this ``susceptibility''. We probably need to return to this after introducing the model. Also, we can compare this with Amelkin's works on polar dynamics.}
We will stack the susceptibility values of all the agents in a vector $\mathbf z$ to be referred to as the susceptibility profile. We shall denote by $\mathcal Z=(0,+\infty)^n$ the set of susceptibility profiles. 

The agents all observe noisy versions of a common state of the world $\theta$ in $\mathbb R$. Specifically, every agent $i$ in $\mathcal V$ observes a noisy quantity $$x_i(0)=\theta + \xi_i$$ where $\xi_i$ is a zero-mean random variable with positive finite variance $\sigma^2_i$ in $(0,+\infty)$. We assume that $\xi_i$ and $\xi_j$ are uncorrelated for all~${i \neq j}$.

Aiming to achieve a better estimate than what they initially observed, the agents enter a discussion with their neighbors in $\mathcal G$.
For every $t \in \mathbb R^+$ and $i \in \mathcal V$, the opinion evolution is described by the distributed averaging dynamics
\begin{equation}
    \label{Abelson} 
        \dot x_i(t)  =z_i\sum_jW_{ij}(x_j(t)-x_i(t)).
\end{equation}
These dynamics (with~${z_i=1}$ for all~$i$) were first considered by Abelson~\cite{Abelson64} and later rediscovered in control theory as a continuous-time consensus algorithm~\cite{moreau}. In our model, each agent is featured by the susceptibility coefficient $z_i$, determining (along with the influence weights $W_{ij}$) the effective influence rates between $i$ and the other individuals. The smaller $z_i>0$ is, the more reluctant the individual $i$ becomes to opinion change. The case if~${z_i = 0}$, excluded in this work, corresponds to a stubborn individual who retains her initial opinion. 

The following result will be used to motivate the rest of our model, with the proof being delayed until the Appendix. 
\begin{proposition} \label{prop:xconv}
    The Abelson model~\eqref{Abelson} converges, for every initial condition ${x_i(0) = \theta + \xi_i}$ with $\Var[{\xi_i}] = \sigma_i^2$ and susceptibility vector $\mathbf z$ in $\mathcal Z$, to a consensus. Furthermore, the consensus value is a random variable with mean $\theta$ and variance 
    \begin{align}\label{eq:vz} 
        v(\mathbf z) := \Var[x_i(\infty)] 
        = \left (\sum_j \dfrac{\mu_j}{z_j} \right)^{-2} \!\!\! \sum_k \dfrac{{\mu_k}^2 \sigma_k^2}{z_k^2}.
    \end{align}
\end{proposition}

Clearly, the choice of $\mathbf z$ directly affects the consensus value of the system, with different $\mathbf z$ vectors resulting in more or less variance in the final result. Suppose then that we have a system planner wishing to minimize $v(\mathbf z)$. Assuming either knowledge \tcb{or estimation~\cite{You2017-mf}} of every $\mu_i$ and $\sigma_i$, the system planner's best choice is outlined in the following result, proved in the Appendix.

\begin{proposition} \label{prop:sysopt}
    Let $\sigma^2$ be the column vector containing entries~${\sigma^2_1, ..., \sigma^2_n}$. 
    Then, for every susceptivity profile $\mathbf z$ in $\mathcal Z$, 
    \begin{equation}
    v(\mathbf z)\ge\left(\sum_k\frac1{\sigma_k^2}\right)^{-1}
    \end{equation}
    with equality if and only if 
    $$\mathbf z\in\mathcal Z^*:= \left \{ \alpha [\mu] \sigma^2 \; : \; \alpha > 0 \right \}.$$    
\end{proposition}
Proposition \ref{prop:sysopt} states that the minimum of the consensus variance $v(\mathbf z)$ over all susceptibility profiles $\mathbf z$ in $\mc Z$ is achieved on $\mathcal{Z}^*\subset\mathcal{Z}$. The optimal value can be better interpreted in terms of the so-called expertise (or precision), defined as the inverse of the variance. 
    In fact, Proposition \ref{prop:sysopt} implies that, for every $\mathbf z^*$ in~$\mathcal Z^*$, we have 
    $$\frac1{v(\mathbf z^*)}=\sum_k\frac{1}{\sigma_k^2}\,,$$
    i.e., the optimal group's expertise (``collective wisdom'') is the aggregate of the single agents' expertises.

\tcb{The existence of a system planner with knowledge of every $\mu_i$ and $\sigma_i$ is quite a strong assumption, so the question of investigating alternatives presents itself naturally.} What if each agent instead had knowledge of these parameters only in a certain neighborhood, and worked to minimize the variance? To answer this question, we introduce a family of continuous-time susceptibility dynamics as follows. Consider a second social network modeled as a finite directed weighed graph ${\overline{\mathcal G} = (\mathcal V, \overline{\mathcal E}, \overline{W})}$. The agent set is the same as before, the possible difference being the links and their weights, see also Figs.\ref{fig:G}--\ref{fig:Gbar}. 

\begin{assumption}\label{asm:connect}
    The graph $\overline{\mc G}$ is strongly connected and has a self-loop at every node, that is, $\overline W_{ii}>0$. 
\end{assumption}

The assumption that $\bar{\mc G}$ contains self-loops at each node is natural here, as the new graph describes information flow instead of opinion formation. 
The learning policy proposed below assumes that each agent $i$ is aware of her own parameters $\pi_i,\sigma_i$ along with parameters of some neighbors, and these neighbors need not be the individuals who influence agent $i$'s opinion in the system~\eqref{Abelson}.

The agents communicate over the second network with the goal of finding susceptibility parameters that lower the variance of the opinion-dynamical consensus value. Each agent minimizes a cost function corresponding to the variance in her neighborhood using the following distributed learning policy 
\begin{subequations}
\begin{align} 
        z_i(0)              &> 0, \label{eq:zinitial} \\
        \dot z_i(t)         &= -\dfrac{\partial u_i}{\partial z_i}(\mathbf z(t)), \label{eq:zdyn} \\
        u_i(\mathbf z)   &= \left (\sum_{j} \overline W_{ij}\dfrac{\mu_j}{z_j} \right)^{-2} \!\!\! \sum_{k} \overline W_{ik}\dfrac{{ \mu_k}^2 \sigma_k^2}{z_k^2}. \label{eq:udef}
\end{align}
\end{subequations}
The cost functions $u_i$ are local analogues to the global variance \tcb{function~(\ref{eq:vz})}. The centralities~$\mu_i$ refer to the social network~$\mathcal G$, providing information on the structure of the opinion model to the susceptibility dynamics. The following results are the main contribution of this paper.

\begin{proposition} \label{prop:fixpts}
    The set of optimal susceptibility vectors $\mc Z^*$ exactly comprises the positive equilibria of system~(\ref{eq:zdyn}).
\end{proposition}

\begin{theorem}\label{thm:main}
     For every initial condition $z(0)>0$, the solution to system~(\ref{eq:zdyn}) exists, is unique, and \tcb{remains strictly positive} (i.e., $z(t)>0$ for all $t>0$). Moreover, every such solution converges to a point in $\mc Z^*$.          
\end{theorem}

Specifically, this means that our distributed algorithm manages to find the system optimal solution previously mentioned in Proposition~\ref{prop:sysopt}, without each agent having global knowledge of every $\mu_i$ and $\sigma_i$. A knowledge of these parameters in some local neighborhood is sufficient, as long as the underlying graph is strongly connected. Section~\ref{sec:analysis} is devoted to proving all of the components of this result.

\section{Analysis} \label{sec:analysis}

The analysis of the distributed learning policy~(\ref{eq:zdyn}) is based on a change of variables that effectively reduces it to a non-linear consensus algorithm over a static graph. Specifically, we introduce the substitution $\mathbf{z}\mapsto\mathbf{y}$, where
\begin{equation}\label{eq.var-change}
y_i := \frac{\mu_i \sigma_i^2}{z_i}, \quad i \in \mathcal{V}.
\end{equation}
Obviously, $y_i>0$ if and only if $z_i>0$.

\subsection*{The learning dynamics in the new variables}

We first express the utility function~(\ref{eq:udef}) in the new variables:
\begin{align*}
    u_i( \mathbf y) = \dfrac{A_i(\mathbf y)}{ B_i^2(\mathbf y)},
\end{align*}
where the functions $A_i(\mathbf{y})$ and $B_i(\mathbf{y})$ are defined as
\begin{align*}
    A_i (\mathbf y) := \sum_j \overline{W}_{ij} \dfrac{y_j^2}{\sigma_j^2},\;\; 
    B_i (\mathbf y) := \sum_j \overline{W}_{ij} \dfrac{y_j}{\sigma_j^2}\quad\forall i\in\mathcal{V}. 
\end{align*}
A straightforward computation yields
\begin{equation} \label{eq:yutil}
\begin{aligned}
    \dfrac{\partial u_i}{\partial y_i} &= \dfrac{2 \overline{W}_{ii}}{\sigma_i^2} \dfrac{y_i B_i(\mathbf y) - A_i(\mathbf y)}{B_i^3(\mathbf y)}\\
    &= \dfrac{2 \overline{W}_{ii}}{\sigma_i^2} \dfrac{y_i \sum_j \overline{W}_{ij} \dfrac{y_j}{\sigma_j^2} - \sum_j \overline{W}_{ij} \dfrac{y_j^2}{\sigma_j^2} }{B_i^3(\mathbf y)} \\
    &= \dfrac{2 \overline{W}_{ii}}{\sigma_i^2 B_i^3(\mathbf y)} {\sum_j \overline{W}_{ij} \dfrac{y_j}{\sigma_j^2} (y_i - y_j) }.
\end{aligned}
\end{equation}

By using the chain rule
\begin{align*}
    \dfrac{\partial u_i}{\partial z_i} &= \dfrac{\partial y_i}{\partial z_i} \dfrac{\partial u_i}{\partial y_i} 
    = - \dfrac{\mu_i \sigma_i^2}{z_i^2} \dfrac{\partial u_i}{\partial y_i}=-\frac{y_i^2}{\mu_i\sigma_i^2}\dfrac{\partial u_i}{\partial y_i},
\end{align*}
the learning dynamics~\eqref{eq:zdyn} shapes into
\begin{equation}\label{eq:ydyn}
\begin{aligned}
        \dot y_i &= \dfrac{\partial y_i}{\partial z_i} \dot z_i=-\left(\dfrac{\partial y_i}{\partial z_i}\right)^2\dfrac{\partial u_i}{\partial y_i}=-\dfrac{y_i^4}{\mu_i^2\sigma_i^4}\dfrac{\partial u_i}{\partial y_i}\\
        %&= \dfrac{y_i^2}{\mu_i \sigma_i^2} \dfrac{\dfrac{2 \overline{W}_{ii} y_i^2}{\mu_i \sigma_i^4} \left [ \sum_j \overline{W}_{ij} \dfrac{y_j^2}{\sigma_j^2} - {y_i} \sum_j \overline{W}_{ij} \dfrac{y_j}{\sigma_j^2}\right ]}{\left(\sum_k \overline{W}_{ik} \dfrac{y_k}{\sigma_k^2} \right)^3} \\
        %&= \dfrac{2 y_i^4 \overline{W}_{ii}}{\mu_i^2 \sigma_i^6} \dfrac{ \sum_j \dfrac{\overline{W}_{ij} y_j}{\sigma_j^2} (y_j - y_i) }{\left(\sum_k \overline{W}_{ik} \dfrac{y_k}{\sigma_k^2} \right)^3} \\
        &=\dfrac{2 y_i^4 \overline{W}_{ii}}{\mu_i^2 \sigma_i^6B_i(\mathbf y)^3} \sum_j \dfrac{\overline{W}_{ij} y_j}{\sigma_j^2} (y_j - y_i)\\
        &= \sum_j m_{ij}(\mathbf y) \, (y_j - y_i).
    \end{aligned}
\end{equation}
Here the nonlinear coupling functions are found as
\begin{equation}\label{eq:coupling}
        m_{ij} (\mathbf y) := \dfrac{2 y_i^4 y_j \overline{W}_{ii} \overline{W}_{ij}}{\mu_i^2 \sigma_i^6 \sigma_j^2}  \left(\sum_k \overline{W}_{ik} \dfrac{y_k}{\sigma_k^2} \right)^{-3} \; \forall i, j \in \mathcal V.
\end{equation}
Notice that, for every $\mathbf{y}>0$, the coupling weight $m_{ij}(\mathbf{y})$ is positive if and only if
$\bar W_{ij}>0$.

\subsection*{The proof of Proposition~\ref{prop:fixpts}}

Notice first that if $\mathbf{z}^* \in \mc{Z}^*$, then by definition there exists $\alpha > 0$ such that
\[
\frac{\mu_1 \sigma_1^2}{z_1^*} = \cdots = \frac{\mu_n \sigma_n^2}{z_n^*} = \frac{1}{\alpha}.
\]
Denoting by $y^*$ the new coordinates corresponding to $z^*$, one proves that $\mathbf z^* \in \mc Z^*$ if and only if $\mathbf y^*$ is a consensus vector, that is, $\mathbf y^* =c\mathbb{1}$ for some $c>0$. 

Proposition~\ref{prop:fixpts} can be restated as follows: a vector $\mathbf{y}^*>0$ is an equilibrium of~\eqref{eq:ydyn} if and only if it is a consensus vector. This result follows directly from Assumption~\ref{asm:connect} and~\eqref{eq:coupling}. Indeed, the equations~\eqref{eq:ydyn} can be written in compact form as
\[
\dot{\bf y}=-\mathcal{L}[M(\mathbf{y})]y,
\]
where the Laplacian matrix corresponds to the matrix $M(\mathbf{y})=(m_{ij}(\mathbf{y}))$.
Hence, for every $\mathbf{y}>0$, the graph associated with $\mathbf{L}(\mathbf{y})$ inherits its strong connectivity from the graph of matrix $\overline W$. Hence, $\mathbf{y}^*$ is an equilibrium if and only if $\mathbf{y}^*\in\ker \mathbf{L(y^*)}$, and this kernel is coincident with the set of all consensus vectors.

\subsection*{The proof of Theorem~\ref{thm:main}}

Notice that as long as the solution $\mathbf{y}(t)$ remains strictly positive, the dynamical system~\eqref{eq:ydyn} constitutes a special case of the nonlinear averaging dynamics examined in~\cite{Lin2007}.
Using the notation from~\cite{Lin2007}, our case corresponds to a time-invariant interaction digraph, denoted by 
$\mathcal{\bar G}$, and the scalar states of the nodes $y_i\in\mathbb{R}$, satisfying the time-invariant dynamics
\[
\dot y_i=f^i(\mathbf{y})=\sum_j m_{ij}(\mathbf y) \, (y_j - y_i).
\]
Obviously, $a_{ij}(\mathbf y)$ are positive and locally Lipschitz in $\mathcal{Z}=(0,\infty)^n$. Theorem~3.6 in~\cite{Lin2007} implies that if a closed convex set $\mathcal{A}$ contains $y_i(0)$ for every $i\in\mathcal{V}$, then it contains every coordinate $y_i(t)$ of the solution\footnote{In~\cite{Lin2007}, such a set $\mathcal{A}$ is referred to as “positively invariant”; in ODE theory, one typically states that the subset of the state space $\mathcal{A}^n$ is forward invariant.} for all $t \geq 0$. Applying this result with $\mathcal{A} = [\min_i y_i(0), \max_i y_i(0)]$, we deduce that the solution $\mathbf{y}(t)$ remains uniformly positive, bounded, and globally defined for all time. Reverting to the original variables $\mathbf{z}$, this implies that each component remains positive and upper bounded, ensuring that the solution exists for all $t \geq 0$.

Applying Theorem~3.8 from~\cite{Lin2007} to the closed and convex set
$\mathcal{S} = [\min_i y_i(0), \max_i y_i(0)]$
yields that the solution $\mathbf{y}(t)$ exhibits asymptotic agreement:
\begin{equation}\label{eq:asympt-agr}
\lim_{t\to\infty}|y_i(t)-y_j(t)|=0\quad\forall i,j\in\mathcal{V}.    
\end{equation}
In fact, this implies a stronger fact: each $y_i(t)$ converges to a common limit. Indeed, since the function $\mathbf{y}(t)$ is uniformly bounded, a sequence $t_k\to +\infty$ exists such that $\mathbf{y}(t_k)$ converges; \eqref{eq:asympt-agr} then guarantees that the limit must be the consensus vector:
$\mathbf{y}(t_k)\xrightarrow[k\to\infty]{} \mathbb{y}^*=\zeta\mathbb{1}$ for $\zeta\in\mathbb{R}$. Moreover,~\cite[Theorem~3.7]{Lin2007} implies that consensus vectors are Lyapunov (weakly) stable equilibria of~\eqref{eq:ydyn}. That is, for every $\varepsilon>0$ there exists $\delta\in(0,\varepsilon)$ such that, once the solution $\mathbf{y}(t)$ enters the open $\delta$-ball $B_{\delta}(\mathbb{y}^*):=\{\mathbf{y}:\|\mathbf{y}-\mathbb{y}^*\|\}<\delta$, it remains in the open $\varepsilon$-ball 
$B_{\varepsilon}(\mathbb{y}^*)$. Hence, for $k$ being so large that $\mathbf{y}(t_k)\in B_{\delta}(\zeta\mathbb{y}^*)$
and $t>t_k$, the solution $\mathbf{y}(t)$ stays in $B_{\varepsilon}(\mathbb{y}^*)$.
Since $\varepsilon>0$ can be arbitrary, we prove that, in reality, $\mathbf{y}(t)\xrightarrow[t\to\infty]{} \mathbb{y}^*=\zeta\mathbb{1}$. 
Recalling that $\mathbf{y}(t)$ is uniformly positive, we have $\zeta>0$.
Returning to the variables $\mathbf{z}$, one has
\[
z_i(t)\xrightarrow[t\to\infty]{} z_i^*:=\dfrac{\mu_i\sigma_i^2}{\zeta},
\]
and the ultimate vector of susceptibilities $z^*$ belongs to $\mathcal{Z}^*$.

\begin{remark}
    While the limit $z^*=\lim_{t\to\infty}\mathbf{z}(t)$ is not explicitly known, the proof of Theorem~\ref{thm:main}
    implies that $\min_i y_i(0)\leq y_i^*=\mu_i \sigma_i^2/z_i^*\leq\max_i y_i(0)$. This gives a simple yet conservative estimate of the fixed point found by the distributed algorithm.
\end{remark}

\section{Numerical Example} \label{sec:num}

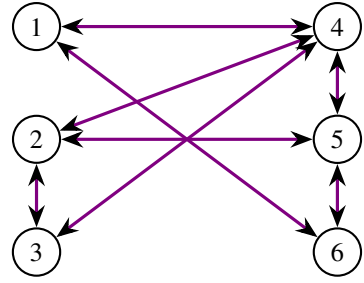
\begin{figure}
    \centering
    
    \begin{tikzpicture}
    \begin{scope}[every node/.style={circle,thick,draw}]
        \node (1) at (0,3) {1};
        \node (2) at (0,1.5) {2};
        \node (3) at (0,0) {3};
        \node (4) at (4,3) {4};
        \node (5) at (4,1.5) {5};
        \node (6) at (4,0) {6} ;
    \end{scope}
    
    \begin{scope}[>={Stealth[black]},
                  every node/.style={fill=white,circle},
                  every edge/.style={draw=violet,very thick}]
                  
        \path [<->] (1) edge (4);
        \path [<->] (1) edge (6);
        
        \path [<->] (2) edge (3);
        \path [<->] (2) edge (4);
        \path [<->] (2) edge (5);

        \path [<->] (3) edge (4);
        
        \path [<->] (4) edge (5);
        
        \path [<->] (5) edge (6);
    \end{scope}
    \end{tikzpicture}

    \caption{The social network graph $\mathcal G$}
    \label{fig:G}
\end{figure}

To illustrate the behavior of our dynamical system, we construct\footnote{All simulations were computed in MATLAB using the ode45 and ode23s solvers. Code is available on request from the first author.} a network of ${n=6}$ agents connected as per Figure~\ref{fig:G}, with all edge weights equal to 1. The vector of agents' variances $\sigma^2$ is set to $[1 , 1.1 , 1 , 1.2 , 1.1 , 1 ]$ and the centrality vector $\mu=\left [1/8, 3/16, 1/8, 1/4, 3/16, 1/8 \right]$. Agent 4 is the most central, but they also have the highest variance on their measurement of the ground truth. If susceptibilities were uniformly distributed, the noisy agent would have more impact on the steady-state value compared to the agents with lower variances.

\begin{figure}
    \centering
    
    \begin{tikzpicture}
    \begin{scope}[every node/.style={circle,thick,draw}]
        \node (1) at (-1,2.4) {1};
        \node (2) at (-1,0) {2};
        \node (3) at (1,1.2) {3};
        \node (4) at (3,1.2) {4};
        \node (5) at (5,2.4) {5};
        \node (6) at (5,0) {6} ;
    \end{scope}
    
    \begin{scope}[>={Stealth[black]},
                  every node/.style={fill=white,circle},
                  every edge/.style={draw=blue,very thick}]
                  
        \path [<->] (1) edge (2);
        \path [<->] (1) edge (3);
        
        \path [<->] (2) edge (3);

        \path [<->] (3) edge (4);
        
        \path [<->] (4) edge (5);
        \path [<->] (4) edge (6);
        
        \path [<->] (5) edge (6);
    \end{scope}
    \end{tikzpicture}

    \caption{The susceptibility optimization graph $\overline{\mathcal G}$}
    \label{fig:Gbar}
\end{figure}
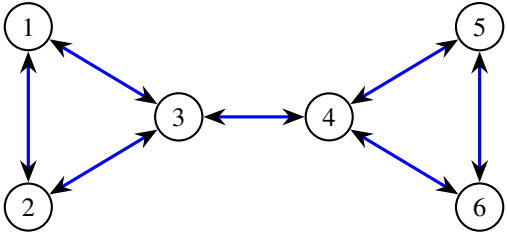

The graph $\overline{\mc G}$ used for susceptibility optimization is shown in Figure~\ref{fig:Gbar}, with all link weights once again being set to 1. The bar graph is an interesting example for our model as one would \tcb{intuitively} expect the bottleneck between nodes~3 and~4 to affect an algorithm attempting to find the globally optimal solution. Nevertheless, due to the dynamics essentially being \tcb{an averaging system, the performance is optimal meaning that the variance is minimized. The scalability of such algorithms depends on local complexity of the graph, since every agent only sees their own neighborhood. Therefore, large graphs do not present a problem as long as they are sparse.}

\begin{figure}[!ht]
    \centering
    \includegraphics[width=\linewidth]{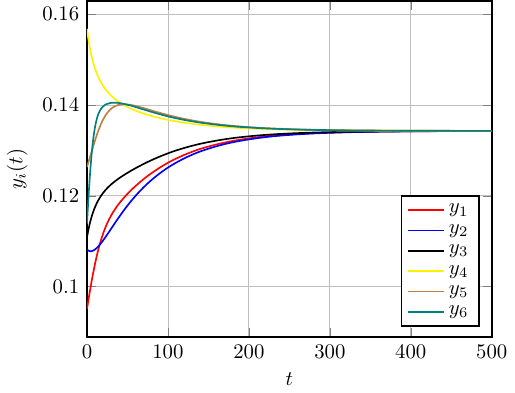}        
    % \begin{tikzpicture}
    %     \begin{axis}[
    %         xmin=0, xmax=500,
    %         xlabel={$t$}, 
    %         %ylabel={$\mu_i \sigma_i^2 / z_i(t)$},
    %         ylabel={$y_i(t)$},
    %         legend pos=south east,
    %         grid=major,
    %         cycle list name=color list,
    %         thick
    %     ]
    %         \addplot table [
    %                 x index=6, 
    %                 y index=0, 
    %                 col sep=comma] 
    %                 {barbell.txt}; 
    %         \addplot table [
    %                 x index=6, 
    %                 y index=1, 
    %                 col sep=comma] 
    %                 {barbell.txt}; 
    %         \addplot table [
    %                 x index=6, 
    %                 y index=2, 
    %                 col sep=comma] 
    %                 {barbell.txt}; 
    %         \addplot table [
    %                 x index=6, 
    %                 y index=3, 
    %                 col sep=comma] 
    %                 {barbell.txt}; 
    %         \addplot table [
    %                 x index=6, 
    %                 y index=4, 
    %                 col sep=comma] 
    %                 {barbell.txt}; 
    %         \addplot table [
    %                 x index=6, 
    %                 y index=5, 
    %                 col sep=comma] 
    %                 {barbell.txt}; 
        
    %     \legend{$y_1$, $y_2$, $y_3$, $y_4$, $y_5$, $y_6$};
    %     % \legend{
    %     % $\mu_1 \sigma_1^2 / z_1$,
    %     % $\mu_2 \sigma_2^2 / z_2$,
    %     % $\mu_3 \sigma_3^2 / z_3$,
    %     % $\mu_4 \sigma_4^2 / z_4$,
    %     % $\mu_5 \sigma_5^2 / z_5$,
    %     % $\mu_6 \sigma_6^2 / z_6$
    %     % };
    %     \end{axis}
    % \end{tikzpicture}
    
    \caption{In the variable ${y_i (t) = \mu_i \sigma_i^2 / z_i(t)}$ the agents reach consensus. This serves as an illustration of the fact that an optimal susceptibility configuration is an egalitarian one where each agent listens to others in direct proportion to their own variance and their own centrality.}
    \label{fig:simulationY}
\end{figure}

Figure~\ref{fig:simulationY} \tcb{shows} the dynamics in the $y_i$ variables. There, the system converges to a consensus. This is \tcb{an illustration} of the character of the points in the optimal set - they are the ones where each agent's susceptibility to change is proportional to its variance and its centrality. Both of these observations make intuitive sense. Agents with high variances benefit from taking other opinions into account, reducing their variance. Those with high centrality contribute disproportionately to the consensus value and should also mediate this \tcb{effect} through their susceptibility.

\begin{figure}[!ht]
    \centering
    \includegraphics[width=\linewidth]{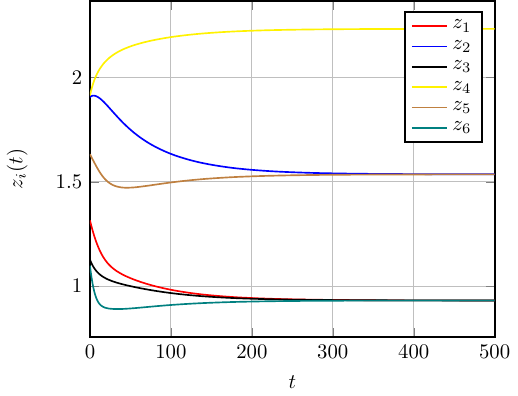}

    % \begin{tikzpicture}
    %     \begin{axis}[
    %         xmin=0, xmax=500,
    %         xlabel={$t$}, ylabel={$z_i(t)$},
    %         legend pos=north east,
    %         grid=major,
    %         cycle list name=color list,
    %         thick
    %     ]
    %         \addplot table [
    %                 x index=6, 
    %                 y index=0, 
    %                 col sep=comma] 
    %                 {barbell_z.txt}; 
    %         \addplot table [
    %                 x index=6, 
    %                 y index=1, 
    %                 col sep=comma] 
    %                 {barbell_z.txt}; 
    %         \addplot table [
    %                 x index=6, 
    %                 y index=2, 
    %                 col sep=comma] 
    %                 {barbell_z.txt}; 
    %         \addplot table [
    %                 x index=6, 
    %                 y index=3, 
    %                 col sep=comma] 
    %                 {barbell_z.txt}; 
    %         \addplot table [
    %                 x index=6, 
    %                 y index=4, 
    %                 col sep=comma] 
    %                 {barbell_z.txt}; 
    %         \addplot table [
    %                 x index=6, 
    %                 y index=5, 
    %                 col sep=comma] 
    %                 {barbell_z.txt}; 
        
    %     \legend{$z_1$, $z_2$, $z_3$, $z_4$, $z_5$, $z_6$};
    %     \end{axis}
    % \end{tikzpicture}

    \caption{Quantities $z_i(t)$ for each agent do not necessarily reach consensus. Here, the agents converge to a dissensus with three groups. The groups form when agents have matching values for both $\mu_i$ and $\sigma_i^2$.}
    \label{fig:simulationZ} 
\end{figure}

Furthermore, the simulation hints at the decreasing convex hull of states mentioned in the proof of Theorem~\ref{thm:main}, as one can clearly see that ${\max_{i \in \mc V} y_i}$ is monotonically decreasing whilst ${\min_{i \in \mc V} y_i}$ is monotonically increasing throughout the dynamics. The exact same dynamics are examined in Figure~\ref{fig:simulationZ}, using the original variables $z_i$ from Equation~\ref{eq.var-change}. This system no longer necessarily converges to a consensus. Instead, a dissensus is the common end result and groups are only formed in situations where agents match each other in both centrality and their individual opinion variance.

\section{Conclusions} \label{sec:conc}
We have proposed a distributed continuous-time dynamical system in which agents learn the optimal variance reducing susceptibilities for the Abelson model. By linking our work to the non-linear consensus literature, we provide analytical results showing that the fixed points of the system are exactly equal to the globally optimal susceptibilities, and that the system converges to a point from the optimal set for all relevant initial conditions. Finally, our numerical example serves to provide intuition on the proof methods by examining the state evolution in different coordinates. 

There exist several possible directions for future work. First, one could let the noise variables $\xi_i$ affecting the initial guesses be correlated. This is a small change making a big difference to the analysis and significantly complicating the utility functions~$u_i$. One could say it would make the modelling scenario more realistic, as e.g. geographically close agents tend to be affected by the same type of environmental, educational or informational influences. Second, one could further generalize the problem by considering also the optimal graph weights for reducing variance and not just the susceptibilities. Finally, a direction that the authors are already pursuing is the analysis of a system where the opinion dynamics evolve in parallel to the susceptibility dynamics. Due to different convergence speeds of the two system components $\mathbf x$ and $\mathbf z$, the analysis is far from trivial as the opinion dynamics do not in general converge to the optimal consensus value even if the susceptibility components do still converge to the previously defined set $\mc Z^*$. We hope to further analyze this issue by introducing separate timescales for the two systems. 

\bibliographystyle{IEEEtran}
\bibliography{references}

\newpage
\appendix 

\subsection{Proof of Proposition~\ref{prop:xconv}}

% \color{violet}
It is known that the Abelson model~\eqref{Abelson} converges to consensus whenever the graph corresponding to the Laplacian matrix $\mathcal{L}([\mathbf z]W)$ is strongly connected\footnote{In reality, consensus in the Abelson model requires only quasi-strongly connectivity (the existence of a directed spanning tree)~\cite{tutp1}, whereas strong connectivity is used in this paper to guarantee that $\mu_i>0$.}~\cite{tutp1,FB-LNS}. This condition is guaranteed by Assumption~\ref{asm:connect}, recalling that $\mathbf{z}>0$. The vector of (identical) final opinions can be found as follows
\begin{align*}
    \lim_{t \rightarrow \infty } \mathbf x(t) = (\mathbf w(\mathbf{z})^\top \mathbf x(0))\mathbb{1},
\end{align*}
where $\mathbf w(\mathbf z)$ is the normalized left eigenvector of the Laplacian $\mathcal{L}([\mathbf z]W)$, corresponding to the eigenvalue $0$. One can show that $\mathbf{w}(\mathbf z)$ is proportional to the vector $[\mathbf z]^{-1}\mu$, that is,
\[
\mathbf{w}(\mathbf{z})=\dfrac{1}{\sum_j \mu_j/z_j} [\mathbf z]^{-1}  \mu.
\]
Hence
the final opinion of each agent is
%\color{black}
\begin{align*} 
    \lim_{t \rightarrow \infty} x_i(t) 
    &= \dfrac{1}{\sum_j \dfrac{ \mu_j}{z_j}} \left([\mathbf z]^{-1}  \mu \right)^\top (\theta \mathbb{1} + \xi) \\
    %&= \dfrac{1}{\sum_j \dfrac{\overline \mu_j}{z_j}} \left({\overline \mu}^\top[\mathbf z]^{-1}  \right) (\theta \mathbb{1} + \xi) \\
    &= \dfrac{1}{\sum_j \dfrac{ \mu_j}{z_j}}  \left({ \mu}^\top[\mathbf z]^{-1}   \theta \mathbb{1} + { \mu}^\top[\mathbf z]^{-1}  \xi \right ) \\
    &= \theta  + \dfrac{\sum_k \dfrac{ \mu_k}{z_k} \xi_k}{\sum_j \dfrac{ \mu_j}{z_j}}.
\end{align*}
Here $\mu$ is the left eigenvector of $W$ defined in~\eqref{eq.mu}.
From this form it is clear that $x_i$ has the expectation $\theta$ and the claim about the variance follows from the fact that $\xi_i, \xi_j$ are assumed to be uncorrelated for all $i \neq j$.

\subsection{Proof of Proposition~\ref{prop:sysopt}}

Denote
\[
\nu_i:=\frac{\mu_i/z_i}{\sum_{j}\mu_j/z_j}.
\]
Then, obviously, the equalities hold as follows:
\[
\begin{gathered}
\upsilon(\mathbf{z})=\sum\nolimits_j\nu_j^2\sigma_j^2,\\
\sum\nolimits_j(\nu_j\sigma_j)\sigma_j^{-1}=\sum\nolimits_j\nu_j=1.
\end{gathered}
\]
Using the Cauchy-Schwarz inequality, one has
\[
\upsilon(\mathbf{z})\sum\nolimits_j\sigma_j^{-2}=\sum\nolimits_j(\nu_j\sigma_j)^2\sum\nolimits_j\sigma_j^{-2}\geq 1,
\]
which inequality is strict unless the two vectors are parallel: $\mu_i\sigma_i/z_i=\nu_i\sigma_i=c\sigma_{i}^{-1}$ for some scalar $c>0$. Hence, $z_i=\alpha\mu_i\sigma_i^2$ for all $i$, where $\alpha=c^{-1}>0$, i.e., $\mathbf{z}\in\mc Z^*$.

\end{document}